\documentclass[3p,times]{elsarticle}

\usepackage{ecrc}
\usepackage{epstopdf} 

\volume{00}

\firstpage{1}

\journalname{Nuclear Physics A}
\runauth{}

\jid{npa}

\jnltitlelogo{Nuclear Physics A}

\usepackage{amssymb}

\biboptions{square,comma,numbers,sort&compress}

\usepackage[figuresright]{rotating}

\begin{document}

\begin{frontmatter}



\dochead{}

\title{Measurements of electrons from heavy-flavour hadron decays in pp, p-Pb and Pb-Pb collisions with ALICE at the LHC}


\author{\vspace*{-0.5cm}
E. Pereira de Oliveira Filho on behalf of the ALICE Collaboration}

\address{Universidade de S\~{a}o Paulo, Departamento de F\'{i}sica Nuclear\\
Travessa R da rua do Mat\~{a}o 187\\
05508900, S\~{a}o Paulo, Brasil
\vspace*{-1.0cm}}

\begin{abstract}
Heavy-flavour hadrons, i. e. hadrons carrying charm or beauty quarks, are a well-suited probe to study the Quark-Gluon Plasma (QGP) in relativistic heavy-ion collisions. For this reason, measurements of electrons from heavy-flavour hadron decays have been performed in pp, p-Pb and Pb-Pb collisions at the LHC with the ALICE detector. Results for the nuclear modification factors ($R_{\rm{pA}}$ and $R_{\rm{AA}}$) support a final-state energy loss of heavy quarks in central Pb-Pb collisions and, in semi-central collisions a positive elliptic flow coefficient $v_{2}$ of electrons from heavy-flavour hadron decays was observed. Furthermore, a double-ridge structure was observed in the measured two-particle angular correlation distribution, triggered by heavy-flavour decay electrons, in high-multiplicity p-Pb collisions relative to low-multiplicity p-Pb collisions and to pp collisions.
\end{abstract}


\end{frontmatter}


\section{Introduction}
\vspace*{-0.2cm}
The main purpose of study using relativistic heavy-ion collisions at the LHC is to investigate the properties of the Quark-Gluon Plasma (QGP).\\
A good probe suited to study the QGP properties is heavy quarks, i. e. charm and beauty. These are produced in hard partonic scattering processes at the initial stage of the collision. Therefore, they work as self-generated probes of the medium created in these experiments.\\
Heavy quarks lose energy while traversing the medium and, in case of sufficient re-scattering, they may even participate in the collective expansion of the medium. Proton-proton and p-Pb collisions serve as references for Pb-Pb collisions, since an extended QGP phase is not expected to be produced in these collision systems. Particularly, the comparison of measurements in p-Pb with those from Pb-Pb collisions allows the separation between cold nuclear matter (CNM) effects and those due to the presence of an extended, hot, and dense medium.\\
Measurement of electrons from semi-leptonic decay of heavy-flavour hadrons is widely used to study heavy-flavour production in high-energy collisions.\\
Measurements of the production cross section, the nuclear modification factors ($R_{\rm{AA}}$ and $R_{\rm{pA}}$) and the elliptic flow coefficient ($v_{2}$) of heavy-flavour decay electrons were performed with ALICE in pp, p-Pb and Pb-Pb collisions. Furthermore, the two-particle angular correlations triggered by electrons from heavy-flavour hadron decays were evaluated in pp and p-Pb collisions.
\vspace*{-0.5cm}
\section{Heavy-flavour decay electron measurement with ALICE}
\begin{figure}[!ht]
\begin{center}
\includegraphics[scale=0.35]{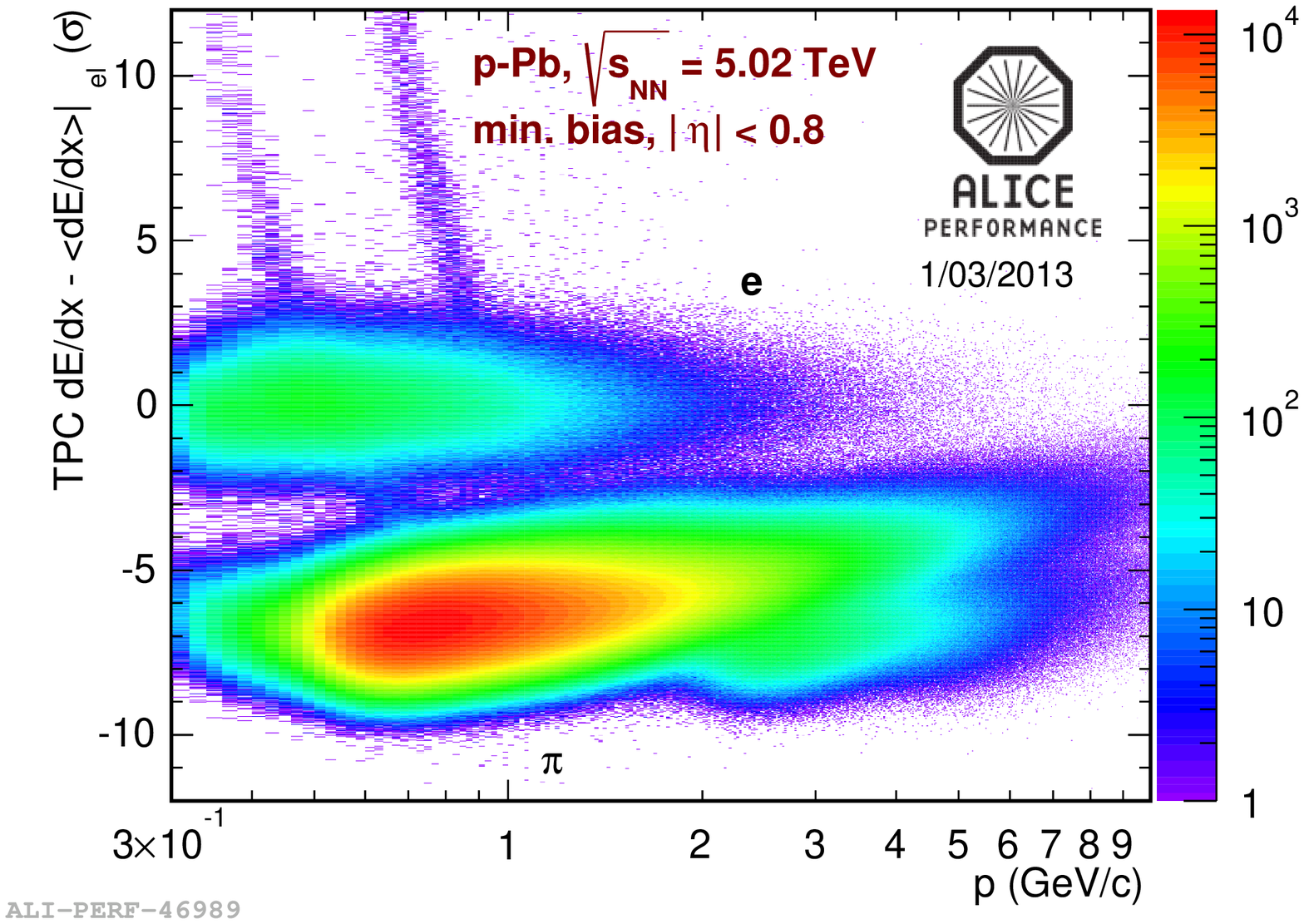}
\includegraphics[scale=0.35]{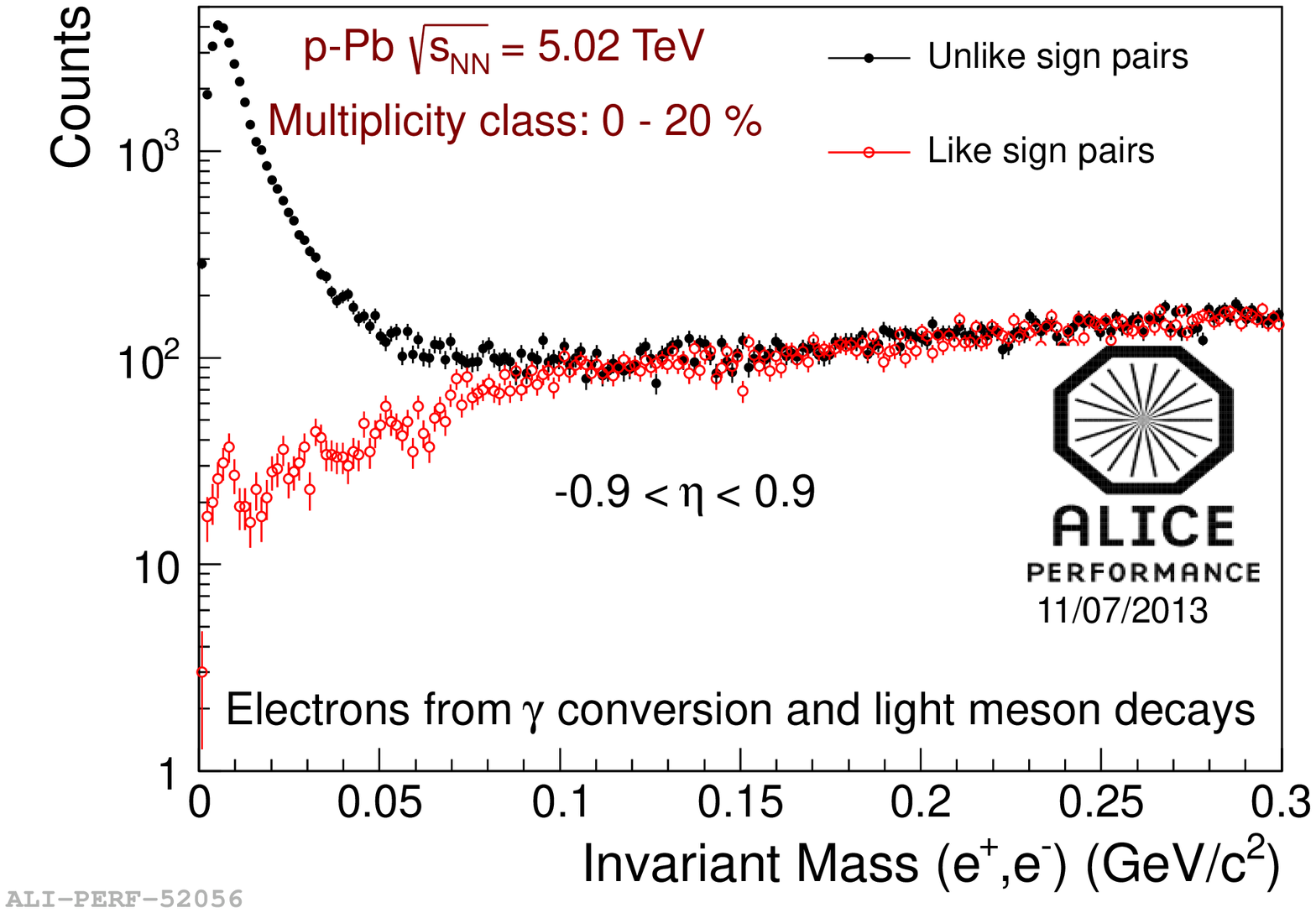}
\caption{Electron identification and background reconstruction in p-Pb collisions at $\sqrt{s_{\rm{NN}}} = 2.76$ TeV. Left panel: The particles' specific energy loss in the TPC relative to the value expected for electrons, in numbers of $\sigma$, as a function of the particle momentum $p$, after requesting the time-of-flight expected for electrons. Right panel: Invariant mass of electron-positron pairs (black filled circles) and of like-sign pairs (red open circles).}
\label{fig:mass}
\end{center}
\end{figure}

\vspace*{-0.2cm}
ALICE (A Large Ion Collider Experiment) is a detector system mainly dedicated to the study of Pb-Pb collisions at the LHC \cite{alice1}. With ALICE, precise three-dimensional track reconstruction and vertex measurements are possible and particle species can be identified over a large momentum range.\\
Electron identification is performed using information of the energy loss ${\rm d}E/{\rm d}x$ in the Time-Projection-Chamber (TPC), the time-of-flight provided by the Time Of Flight (TOF) detector and the $E/p$ ratio measured with the Electromagnetic Calorimeter (EMCal) \cite{alice1, alicehfe}. Figure \ref{fig:mass} (left panel), illustrates the electron selection using the TPC and the TOF detectors.\\
The inclusive electron sample contains not only electrons from heavy-flavour hadron decays but also from various background sources, where the dominant contribution is from Dalitz decays of light mesons, mainly $\pi^{0}$ and $\eta$, and from photon conversions in the detector material. These contributions are statistically subtracted from the electron sample using the two methods described in the following. In the first method, contributions from electron-positron pairs from photon conversions or Dalitz decays were subtracted utilizing the fact that they have small invariant mass. The pair mass distribution is shown in the right panel of Fig. \ref{fig:mass}, where also the combinatorial background, which is estimated via like-sign pairs, is indicated. In the second method, called as cocktail method, the electron background from other sources is calculated using the measured light meson $p_{\rm{T}}$-differential cross section \cite{alicehfe}.
\vspace*{-0.4cm}

\vspace*{-1.2cm}

\section{Nuclear modification factor and elliptic flow}
\begin{figure}[!h]
\begin{center}
\includegraphics[scale=0.47]{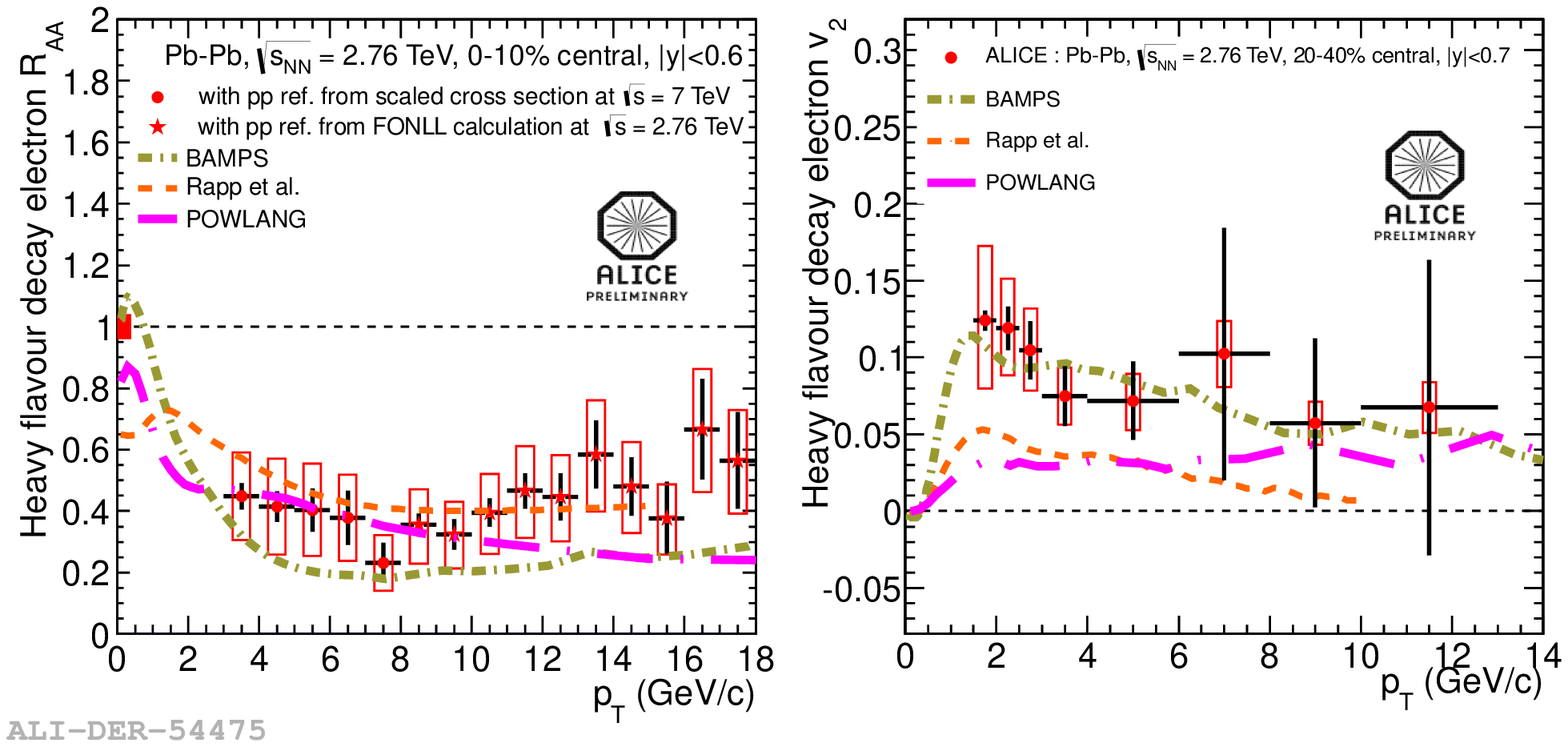}
\includegraphics[scale=0.32]{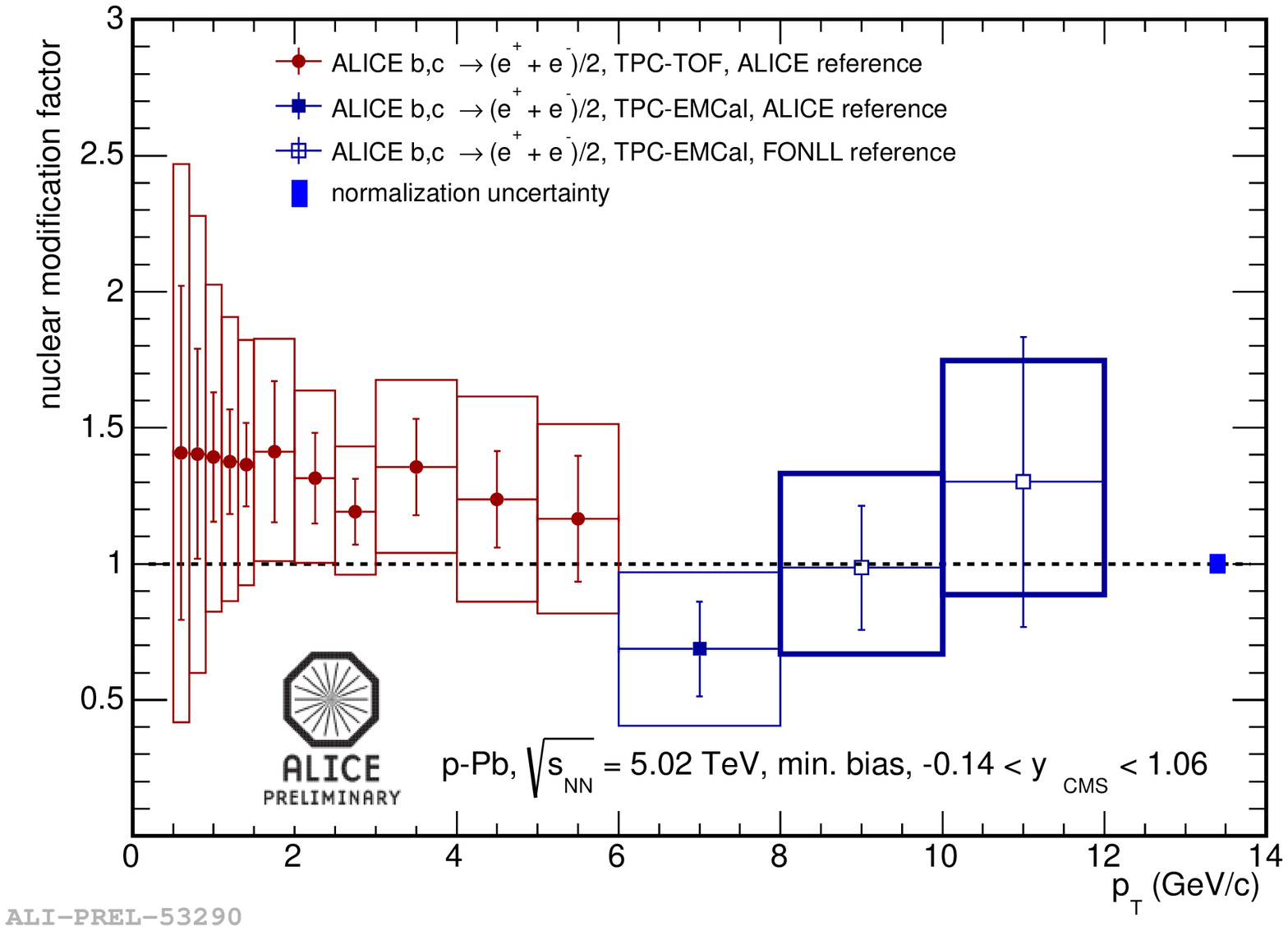}
\caption{Nuclear modification factor (left panel) and elliptic flow coefficient (middle panel) of heavy-flavour decay electrons, as a function of the electron transverse momentum, in Pb-Pb collisions at $\sqrt{s_{\rm{NN}}} = 2.76$ TeV. Several energy-loss models, which are described in references \cite{bamps, rapp, powlang}, are displayed along with the data points. Right panel: Nuclear modification factor of electrons from heavy-flavour hadron decays obtained in p-Pb collisions at $\sqrt{s_{\rm{NN}}} = 5.02$ TeV.}
\label{fig:aa}
\end{center}
\end{figure}

\vspace*{-0.2cm}
The nuclear modification factors ($R_{\rm{AA}}$ and $R_{\rm{pPb}}$) are defined in Equation \ref{eq:raa}, where ${\rm d}N^{\rm{PbPb}} / {\rm d}p_{\rm{T}}$ is the $p_{\rm{T}}$-differential yield in Pb-Pb collisions, ${\rm d}\sigma^{\rm{pp}} / {\rm d}p_{\rm{T}}$ (${\rm d}\sigma^{\rm{pPb}} / {\rm d}p_{\rm{T}}$) is the $p_{\rm{T}}$-differential cross-section in pp (p-Pb) collisions, $\left\langle T_{\rm{PbPb}} \right\rangle$ is the nuclear overlap function in Pb-Pb collisions and $A$ is the mass number of the Pb nucleus. In the absence of medium effects, the nuclear modification factor coincides with unity ($R_{\rm{AA}} = 1$).
\begin{equation}
R_{\rm{PbPb}} = \frac{1}{\left\langle T_{\rm{PbPb}} \right\rangle } \frac{{\rm d}N^{\rm{PbPb}} / {\rm d}p_{\rm{T}}}{{\rm d}\sigma^{\rm{pp}} / {\rm d}p_{\rm{T}}} \qquad ; \qquad R_{\rm{pPb}} = \frac{1}{A} \frac{{\rm d}\sigma^{\rm{pPb}} / {\rm d}p_{\rm{T}}}{{\rm d}\sigma^{\rm{pp}} / {\rm d}p_{\rm{T}}}
\label{eq:raa}
\end{equation}
In Figure \ref{fig:aa} (left panel) the measured $R_{\rm{AA}}$ of heavy-flavour decay electrons is shown as a function of $p_{\rm{T}}$, in central (0-10\%) Pb-Pb collisions at $\sqrt{s_{\rm{NN}}} = 2.76$ TeV. Figure \ref{fig:aa} (right panel) shows the result obtained in p-Pb collisions at $\sqrt{s_{\rm{NN}}} = 5.02$ TeV. A strong suppression of the yield of heavy-flavour decay electrons is observed at high $p_{\rm{T}}$ ($p_{\rm{T}}>3$ GeV/$c$) in central Pb-Pb collisions, relative to the binary scaled cross section in pp collisions, while in p-Pb collisions, the nuclear modification factor ($R_{\rm{pPb}}$) is consistent with unity within statistical and systematic uncertainties. The results from the models calculations are also shown as lines in Fig. \ref{fig:aa}, that include parton energy loss in the hot and dense QCD medium \cite{bamps, rapp, powlang}.\\
The elliptic-flow coefficient $v_{2}$ is defined as $v_{2} = \left\langle \rm{cos \left[ 2\left(\varphi - \Psi_{\rm{RP}} \right) \right]} \right\rangle$, where $\varphi$ is the particle azimuthal angle and $\Psi_{\rm{RP}}$ is the azimuthal angle of the reaction plane. Figure \ref{fig:aa} (middle panel) shows the elliptic-flow coefficient $v_{2}$ of heavy-flavour decay electrons as a function of $p_{\rm{T}}$ in semi-central (20-40\%) Pb-Pb collisions at $\sqrt{s_{\rm{NN}}} = 2.76$ TeV. The elliptic flow of heavy-flavour decay electrons ($v_2^{\rm HFe}$) is deduced by using the Equation \ref{eq:v2}, where $v_{2}^{\rm{incl}}$ and $v_{2}^{\rm{back}}$ are the $v_{2}$ of inclusive and background electrons respectively, and  $R_{\rm{SB}}$ is the signal to background ratio. 
\begin{equation}
\qquad v_{2}^{\rm{HFe}} = \frac{\left(1+R_{\rm{SB}}v_{2}^{\rm{incl}}\right) - v_{2}^{\rm{back}}}{R_{\rm{SB}}}
\label{eq:v2}
\end{equation}
Positive $v_{2}$ for $p_{\rm{T}} < 4$ GeV/$c$ with 3$\sigma$ confidence level might indicate that charm quarks participate in the collective expansion of the QGP. This result is also compared to the aforementioned models. The simultaneous description of the measured $R_{\rm{AA}}$ and $v_{2}$ is challenging for models.
\vspace*{-0.4cm}

\vspace*{-1.2cm}

\section{Two-particle angular correlations}
\begin{figure}[!h]
\begin{center}
\includegraphics[scale=0.35]{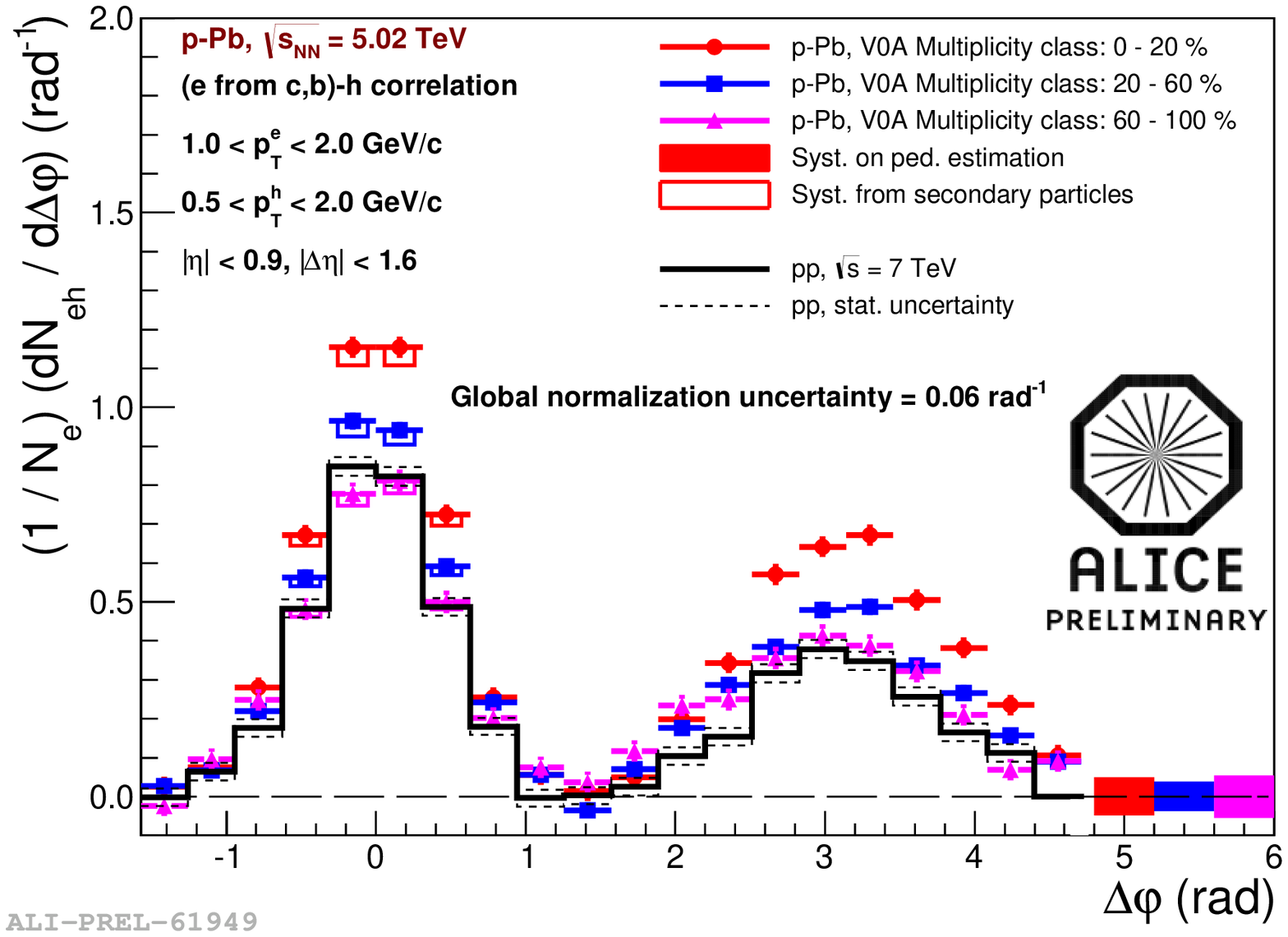}
\includegraphics[scale=0.35]{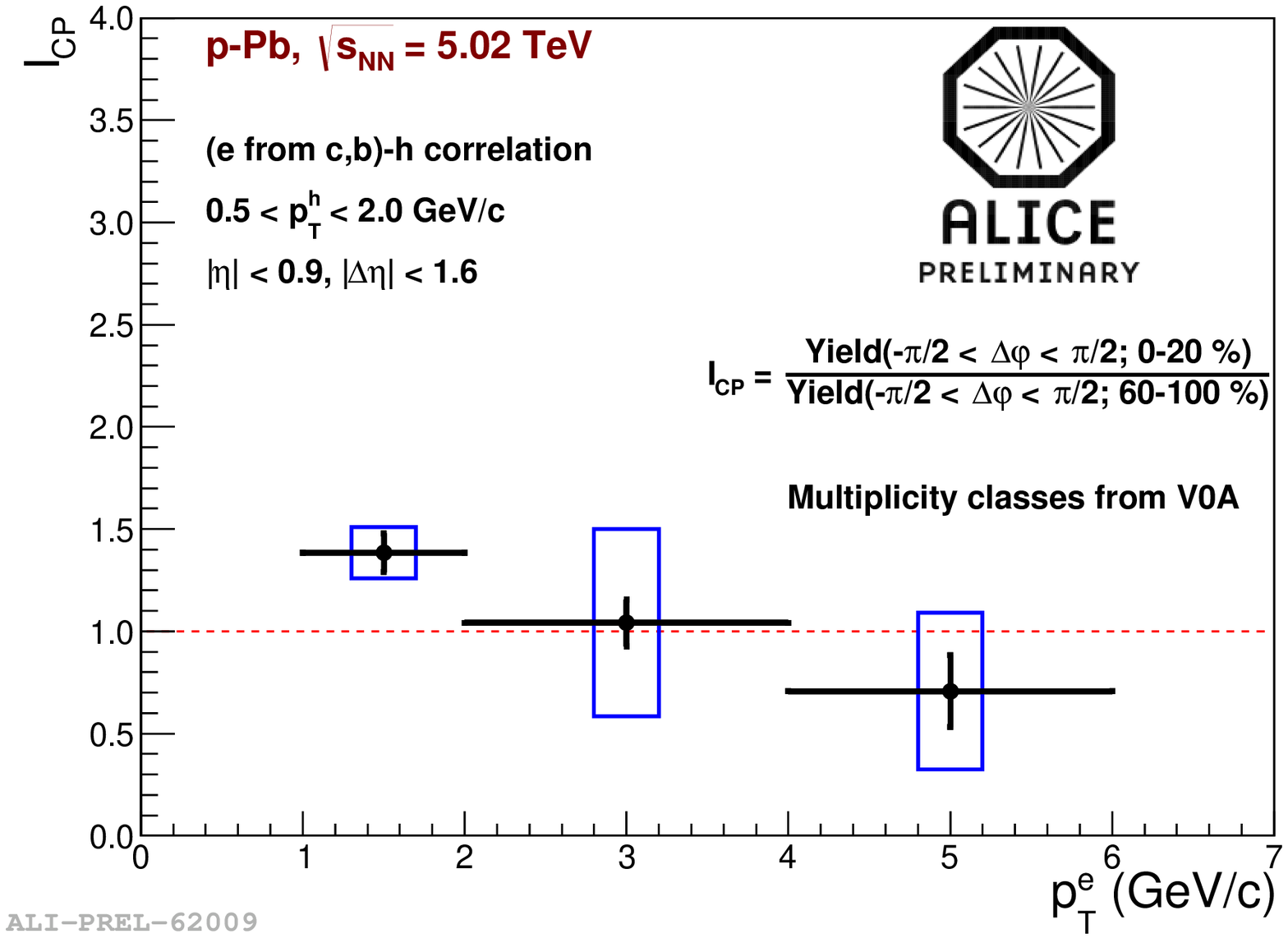}
\caption{Left panel: Azimuthal correlation distribution between heavy-flavour decay electrons and charged particles in high, intermediate and low multiplicity classes in p-Pb collisions at $\sqrt{s_{\rm{NN}}} = 5.02$ TeV and in pp collisions at $\sqrt{s} = 7$ TeV. Right panel: Ratio between the yields in high and low multiplicity p-Pb collisions ($I_{\rm{CP}}$) as a function of the electron $p_{\rm{T}}$.
\vspace*{-0.5cm}}
\label{fig:corr1}
\end{center}
\end{figure}

\begin{figure}[!h]
\begin{center}
\includegraphics[scale=0.35]{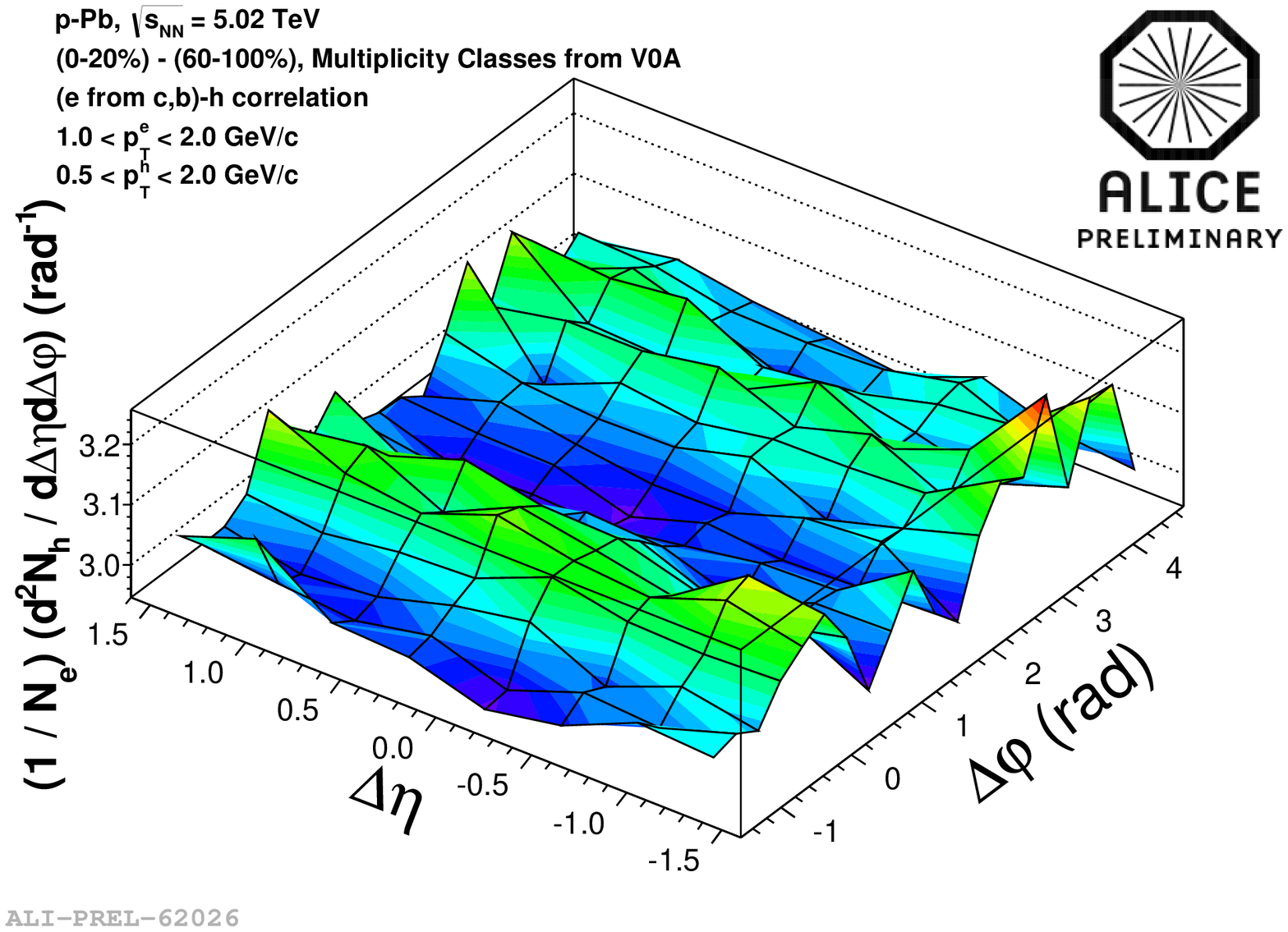}
\includegraphics[scale=0.35]{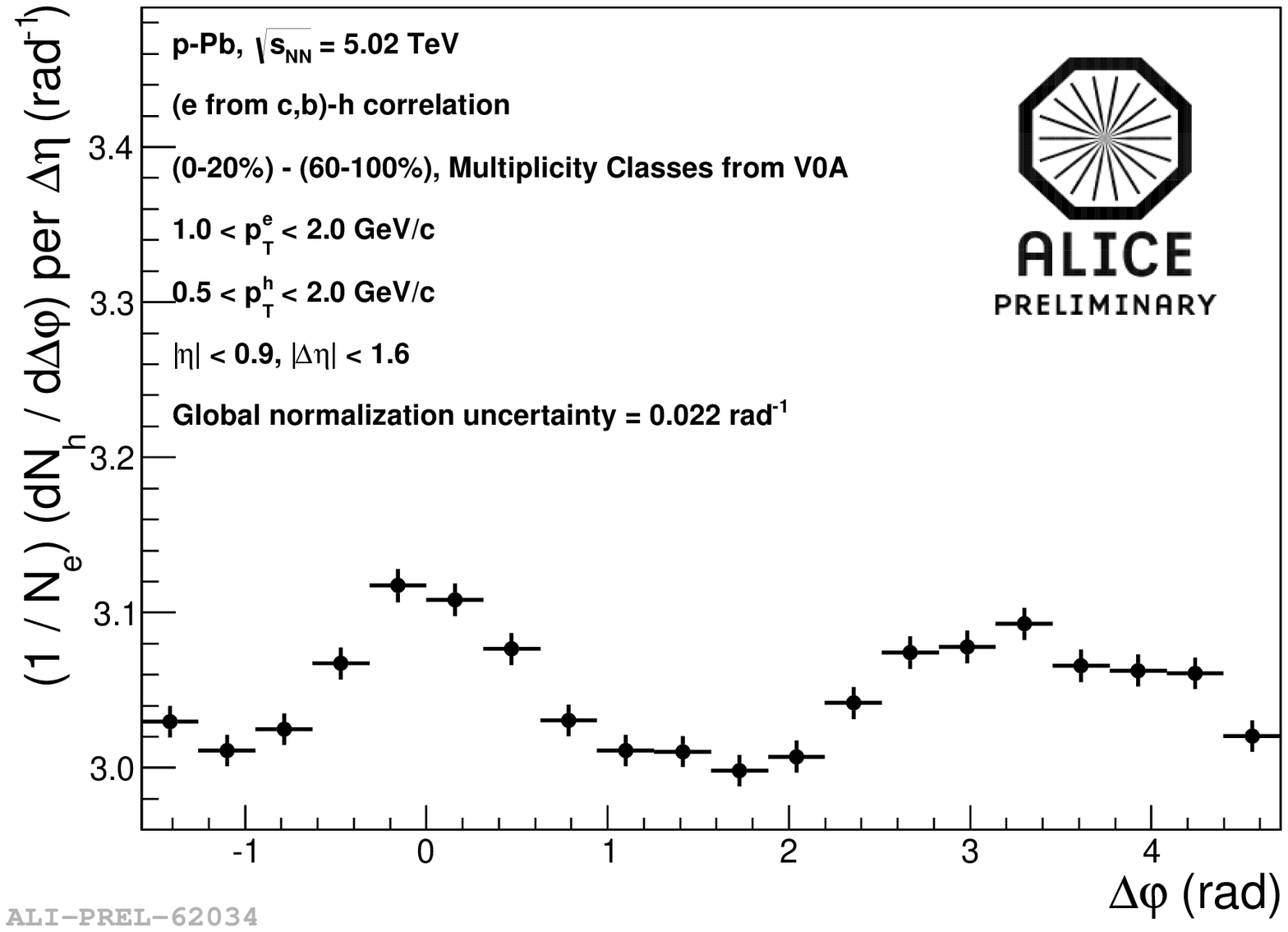}
\caption{Difference between two-particle correlation distribution in high (0-20\%) and low (60-100\%) multiplicity p-Pb collisions, in the ($\Delta\eta$,$\Delta\varphi$) space (left panel) and the projection on the $\Delta\varphi$ axis (right panel).}
\label{fig:corr2}
\end{center}
\end{figure}

\vspace*{-0.2cm}
The two-particle angular correlations between heavy-flavour decay electrons and charged particles produced in the collision were measured in three classes of multiplicity in p-Pb collisions at $\sqrt{s_{\rm{NN}}} = 5.02$ TeV, and in pp collisions at $\sqrt{s} = 7$ TeV. The two-particle correlation distribution is the distribution of the difference in azimuth ($\Delta\varphi$) and in pseudorapidity ($\Delta\eta$) between a trigger (t) and an associated (a) particle, normalized to the number of trigger particles. The correlation distribution is corrected for the pair-finding efficiency obtained using the event mixing technique.\\
In Figure \ref{fig:corr1} (left panel) the measured correlations are presented, after the projection on the $\Delta\varphi$ axis, for three multiplicity classes in p-Pb collisions and for pp collisions, for $1 < p_{\rm{T}}^{e} < 2$ GeV/$c$. The correlation distribution in low-multiplicity (60-100\%) p-Pb collisions is similar to that obtained in pp collisions. However, a modification of the correlation function, on the near ($-\frac{\pi}{2} < \Delta\varphi < \frac{\pi}{2}$) and away ($\frac{\pi}{2} < \Delta\varphi < \frac{3\pi}{2}$) side is visible for electrons at low pT in high-multiplicity p-Pb collisions with respect to pp and low-multiplicity p-Pb collisions. This modification can be quantified by the ratio $I_{\rm{CP}}$ between the per-trigger yields in high (0-20\%) and low (60-100\%) multiplicity p-Pb collisions, as defined in Equation \ref{eq:corr2}.
\vspace*{-0.4cm}
\begin{equation}
I_{\rm{CP}} (p_{\rm{T}}^{e},p_{\rm{T}}^{h}) = \left. \left(\int\limits_{-\frac{\pi}{2}}^{\frac{\pi}{2}} C(\Delta\varphi;p_{\rm{T}}^{e},p_{\rm{T}}^{h}) d(\Delta\varphi)\right)^{0-20 \%}  \middle/ \left(\int\limits_{-\frac{\pi}{2}}^{\frac{\pi}{2}} C(\Delta\varphi;p_{\rm{T}}^{e},p_{\rm{T}}^{h}) d(\Delta\varphi)\right)^{60-100\%} \right.
\label{eq:corr2}
\vspace*{-0.3cm}
\end{equation}
The ratio $I_{\rm{CP}}$ for the near side as a function of the electron $p_{\rm{T}}$ is shown in the right panel of Fig. \ref{fig:corr1}.
\vspace*{-0.4cm}
\newpage
$I_{\rm{CP}}$ is larger than unity for electrons with low $p_{\rm{T}}$. In order to quantify the change of the correlation distribution in the ($\Delta\varphi$,$\Delta\eta$) space, the correlation distribution in low-multiplicity p-Pb collisions was subtracted from the one in high-multiplicity collisions. In Figure \ref{fig:corr2} (left panel) the result of the subtraction is shown in the ($\Delta\eta$,$\Delta\varphi$) space. The double ridge structure displayed in the left panel of Figure \ref{fig:corr2} indicates that the observed modification in the correlation function is mainly due to correlations of long range in pseudorapidity. This is qualitatively similar to what was observed for di-hadron correlation \cite{ALICEpA}. In that case, the observed double-ridge structure can be described by hydrodynamic model calculations assuming an extended system in the final state, as well as by colour glass condensate (CGC) model calculations, through which gluon saturation in the initial state is represented. The responsible mechanism for this structure might affect charm and beauty quarks as well.
\vspace*{-1.2cm}
\section{Conclusions}
\vspace*{-0.2cm}
Measurements of several observables related to heavy-flavour decay electrons were performed with ALICE in pp, p-Pb and Pb-Pb collisions at the LHC. Interesting effects were observed, which can be attributed to the interaction of heavy-quarks with the hot and dense medium produced in Pb-Pb collisions or to the cold nuclear matter effects in p-Pb collisions, respectively.\\ 
In Pb-Pb collisions a strong modification in the $p_{\rm{T}}$-differential yield of electrons from heavy-flavour hadron decays, due to parton energy-loss in the hot and dense QCD medium, was measured. In high-multiplicity p-Pb collisions a double-ridge structure was observed in the two-particle correlation distribution triggered by heavy-flavour decay electrons, similarly to what was observed for light-flavour hadrons, indicating that the responsible mechanism might affect the dynamics of heavy quarks as well.




\vspace*{-0.2cm}
\section*{Acknowledgments}
\vspace*{-0.3cm}
The author thanks the \textit{Conselho Nacional de Desenvolvimento Cient\'ifico e Tecnol\'ogico - CNPq}, for the financial support.
\vspace*{-0.2cm}
\bibliographystyle{elsarticle-num}



\end{document}